\documentclass[aps,prl,superscriptaddress,twocolumn,showpacs]{revtex4}
\usepackage{graphics}
\usepackage{amsmath}
\usepackage{graphicx}
\usepackage{amsfonts}
\usepackage{amssymb}

\begin{document}

\title{Quantum Direct Communication with Continuous Variables}
\author{Stefano Pirandola}
\affiliation{MIT - Research Laboratory of Electronics, Cambridge
MA 02139, USA}
\author{Samuel L. Braunstein}
\affiliation{Computer Science, University of York, York YO10 5DD,
United Kingdom}
\author{Stefano Mancini}
\affiliation{CNISM \& Dipartimento di Fisica, Universit\`a di
Camerino, I-62032 Camerino, Italy}
\author{Seth Lloyd}
\affiliation{MIT - Research Laboratory of Electronics, Cambridge
MA 02139, USA} \affiliation{MIT - Department of Mechanical
Engineering, Cambridge MA 02139, USA}
\date{\today }

\begin{abstract}
We show how continuous variable systems can allow the direct
communication of messages with an acceptable degree of privacy.
This is possible by combining a suitable phase-space encoding of
the plain message with real-time checks of the quantum
communication channel. The resulting protocol works properly when
a small amount of noise affects the quantum channel. If this noise
is non-tolerable, the protocol stops leaving a limited amount of
information to a potential eavesdropper.
\end{abstract}


\maketitle

\section{Introduction}

In recent years, quantum communication protocols have been
extended to the domain of continuous variable (CV) systems, i.e.,
quantum systems, like the bosonic modes of the radiation field,
which are characterized by infinite dimensional Hilbert spaces
\cite{Bra}. In particular, it has been understood how a sender
(Alice) can exploit bosonic modes in order to send analog signals
to a receiver (Bob) and then extract a secret binary key from
these signals \cite{Homo,Hetero}. Beyond the possibility of such a
continuous variable quantum key distribution (QKD), here we show
how to use these systems in order to perform a (quasi)confidential
quantum direct communication (QDC) \cite{Almut}, i.e., the
(quasi)private communication of a message from Alice to Bob which
is directly encoded in CV systems.

The ideal situation for QDC trivially occurs when Alice and Bob
are connected by a noiseless channel. However, in general, this is
not the case and the honest users must randomly switch their
confidential communication with real-time checks on the channel.
As soon as they detect the presence of a non-tolerable noise, they
promptly stop the communication. The maximum noise that can be
tolerated is connected to the maximum amount of information that
they are willing to give up to an eavesdropper. In other words, a
good QDC protocol should enable Alice and Bob to communicate all
the message when the noise is suitably low, while losing a small
amount of information when it is not.

Let us consider a bosonic mode described by quadrature operators $\hat{q}$
and $\hat{p}$, satisfying $[\hat{q},\hat{p}]=i$. An arbitrary state of the
system (density operator $\rho $) must fulfill the uncertainty principle $V(%
\hat{q})V(\hat{p})\geq 1/4$, where $V(\hat{x})=\mathrm{{Tr}(\rho \hat{x}%
^{2})-[{Tr}(\rho \hat{x})]^{2}}$ denotes the variance of the arbitrary
quadrature $\hat{x}=\hat{q}$ or $\hat{p}$. In particular, coherent states
satisfy $V(\hat{q})=V(\hat{p}):=\Delta $, where $\Delta =1/2$ represents the
quantum shot-noise. This is the fundamental noise that affects \emph{%
disjoint }measurements of the quadratures $\hat{q}$ and $\hat{p}$ (homodyne
detection), and it is doubled to $\Delta =1$ when the two quadratures are
\emph{jointly} measured (heterodyne detection). A density operator $\rho $
may be faithfully represented by the Wigner quasi-probability distribution $%
W(q,p)$, whose continuous variables $q$ and $p$ are the eigenvalues of the
quadratures. In this phase-space representation, states with Gaussian Wigner
functions are called \emph{Gaussian states}. This is the case of a coherent
state $|\bar{\alpha}\rangle $, whose Gaussian Wigner function is centered at
$\bar{\alpha}=2^{-1/2}(\bar{q}+i\bar{p})$. For coherent states the detection
of an arbitrary quadrature $\hat{x}$\ provides outcomes $x$ following the
marginal distribution
\begin{equation}
G_{\Delta }(x-\bar{x})=\frac{1}{\sqrt{2\pi \Delta }}\exp \left[ -\frac{(x-%
\bar{x})^{2}}{2\Delta }\right] ~,  \label{Gaussian_Real}
\end{equation}%
where $\Delta =1/2$ for homodyne and $\Delta =1$ for heterodyne.

\section{The protocol}

\begin{figure}[tbph]
\vspace{0cm}
\par
\begin{center}
\includegraphics[width=0.48\textwidth] {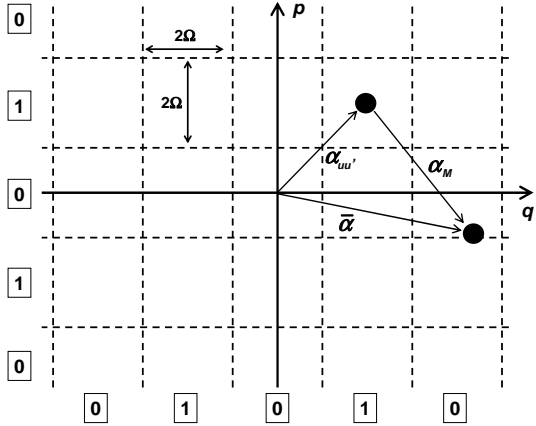}
\end{center}
\par
\vspace{-0.6cm}
\caption{Square lattice in phase-space with unit cell of size $2\Omega $.
Each cell specifies the values of a pair of bits $(u,u^{\prime })$.}
\label{LatticePic}
\end{figure}

Let us show how Alice can transmit message bits by using the phase-space of
a bosonic mode. We discretize the phase-space via a square lattice of
half-step size $\Omega $. Then, an arbitrary cell specifies the values of
two bits $(u,u^{\prime })$ which are given by the parity of its address
along the $q$ and $p$ axes (see Fig.~\ref{LatticePic}). In a simple lattice
encoding, Alice embeds two bits $(u,u^{\prime })$ by \emph{randomly}
choosing a target cell with parities $(u,u^{\prime })$ or, equivalently, by
constructing the message amplitude $\alpha _{uu^{\prime }}$ pointing at the
center of that target cell.
Then, in a first naive protocol, Alice directly prepares the coherent state $%
|\alpha _{uu^{\prime }}\rangle $ from the message amplitude $\alpha
_{uu^{\prime }}$. Such a state is sent to Bob, who performs a heterodyne
detection for extracting $\alpha _{uu^{\prime }}$ and, therefore, the pair $%
(u,u^{\prime })$. Notice that, even in the presence of a noiseless channel,
Bob's decoding cannot be perfect since the Gaussian shape of the coherent
state spreads over the whole phase space and this leads to an \emph{%
intrinsic error}. It is easy to check that the probability of an intrinsic
error (per transmitted bit) is
\begin{equation}
\varepsilon (\Omega ,\Delta )=2\sum_{j=0}^{\infty }\int_{(4j+1)\Omega
}^{(4j+3)\Omega }dx~G_{\Delta }(x)~.  \label{Intrinsic_error}
\end{equation}
In particular, here we fix $\Omega \simeq 2.57$ in order to have
the reasonable low value of $\varepsilon =1\%$. On the one hand,
such a choice for $\Omega $ enables Bob to approach an error-free
decoding when the communication channel is noiseless. On the other
hand, it makes the protocol fragile to eavesdropping since Eve can
optimize her attack to the structure of the lattice, e.g., by
using non-universal quantum cloning machines.

\begin{figure}[tbph]
\vspace{-0.4cm}
\par
\begin{center}
\includegraphics[width=0.49\textwidth] {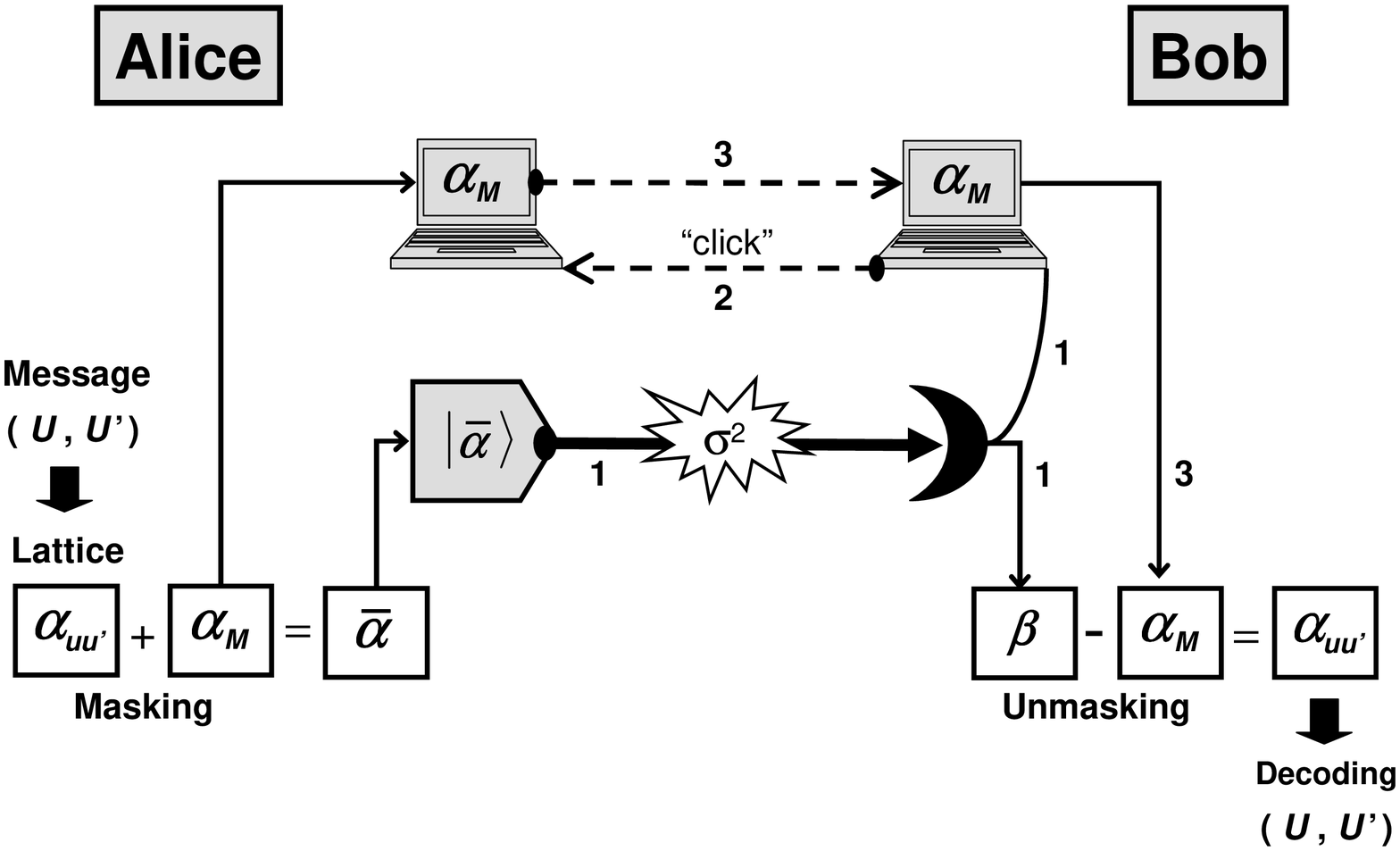}
\end{center}
\par
\vspace{-1.1cm}
\caption{\textbf{Message mode (MM).} \ From the message bits $(u,u^{\prime
}) $, Alice computes the message amplitude $\protect\alpha _{uu^{\prime }}$\
(lattice encoding)\ and then adds the mask $\protect\alpha _{M}$ achieving
the signal amplitude $\bar{\protect\alpha}$. Then, Alice prepares and sends
to Bob the signal state $\left\vert \bar{\protect\alpha}\right\rangle $,
that Bob heterodynes with outcome $\protect\beta $ (step $1$ in the
picture). After detection, Bob classically informs Alice (step $2$) and,
then, Alice classically communicates the mask $\protect\alpha _{M}$ (step $3$%
). At that point, Bob is able to unmask the signal ($\protect\beta -\protect%
\alpha _{M}$), thus reconstructing $\protect\alpha _{uu^{\prime }}$ and,
therefore, $(u,u^{\prime })$.}
\label{MMnoCodesPic}
\end{figure}

Fortunately, we can preclude these strategies by adding a simple classical (%
\emph{masking}) step to the above procedure. In fact, after having computed
the message amplitude $\alpha _{uu^{\prime }}$, Alice can add a \emph{mask}
amplitude $\alpha _{M}$, in such a way that the total \emph{signal}
amplitude $\bar{\alpha}:=\alpha _{M}+\alpha _{uu^{\prime }}$ is continuously
distributed in phase space according to a spread Gaussian (see Fig.~\ref%
{LatticePic}). Then, in a second refined protocol, Alice prepares the
message $\alpha _{uu^{\prime }}$, the mask $\alpha _{M}$ and the signal
state $\left\vert \bar{\alpha}\right\rangle $ (see Fig.~\ref{MMnoCodesPic}).
As a first step, Alice sends the signal state $\left\vert \bar{\alpha}%
\right\rangle $ to Bob, who heterodynes it with outcome $\beta \simeq \bar{%
\alpha}$. Then, after Bob's confirmation of detection, Alice classically
communicates the mask $\alpha _{M}$. As a consequence of these steps, Bob
gets the pair $(\beta ,\alpha _{M})$ from his detection and Alice's
communication. Then, Bob is able to \emph{unmask} the signal by computing $%
\beta -\alpha _{M}\simeq \bar{\alpha}-\alpha _{M}=\alpha _{uu^{\prime }}$
and, therefore, retrieves the\ message bits $(u,u^{\prime })$ via lattice
decoding.
The key-point here is that Eve must choose the probing interaction
before knowing the value of the mask. Since the continuous signal
$\bar{\alpha}$ is highly modulated, the best choice is to adopt a
universal interaction which does not privilege any particular
portion of the phase-space. Here, we consider for Eve the usage of
a universal Gaussian quantum cloning machine (UGQCM)
\cite{GaussCloners}. Such a machine maps the signal state
$\left\vert \bar{\alpha}\right\rangle $ into a pair of output
clones $\rho _{B}$ (sent to Bob)\ and $\rho _{E}$ (taken by Eve),
equal to a Gaussian modulation of $\left\vert
\bar{\alpha}\right\rangle \left\langle \bar{\alpha}\right\vert $
with cloning variances $\sigma _{B}^{2}:=\sigma ^{2}$ and $\sigma
_{E}^{2}=(4\sigma ^{2})^{-1}$. This means that the arbitrary
quadrature $\hat{x}$ of the clone $K=B,E$ has a marginal
distribution equal to $G_{\Delta +\sigma
_{K}^{2}}^{K}(x-\bar{x})$.

\begin{figure}[tbph]
\vspace{-0.6cm}
\par
\begin{center}
\includegraphics[width=0.49\textwidth] {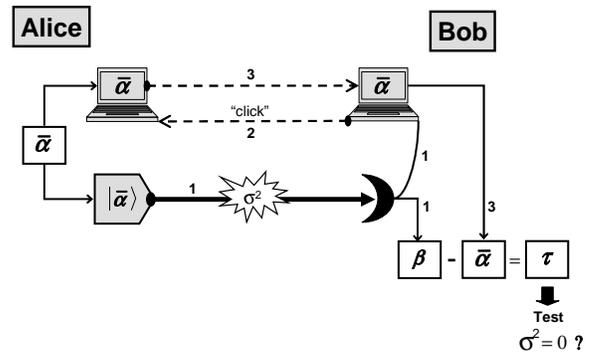}
\end{center}
\par
\vspace{-1.0cm}
\caption{\textbf{Control mode (CM).}\emph{\ }Alice picks up a Gaussian
amplitude $\bar{\protect\alpha}$ and prepares a signal coherent state $%
\left\vert \bar{\protect\alpha}\right\rangle $. Such a state is sent to Bob,
who heterodynes it with outcome $\protect\beta $ (step 1 in the picture).
Then, Bob classically informs Alice (step 2) and Alice communicates the
value of the signal $\bar{\protect\alpha}$ (step 3). At that point, Bob
computes the test variable $\protect\tau :=\protect\beta -\bar{\protect\alpha%
}$ and tests the noise of the channel.}
\label{CMnoCodesPic}
\end{figure}

The above procedure of directly communicating message bits is
called the \emph{message mode} (MM) of the protocol. However,
Alice and Bob have to also perform real-time controls of the added
noise $\sigma ^{2}$ on the channel. This is possible if Alice
randomly switches from message mode instances to suitable
instances of \emph{control mode} (CM) \cite{Ping}. In control mode
(see Fig.~\ref{CMnoCodesPic}), Alice does not process any text
message but only prepares and sends the signal state $\left\vert
\bar{\alpha}\right\rangle $. Then, after
Bob's detection (outcome $\beta $), Alice communicates the value $\bar{\alpha%
}$ of the signal amplitude. At that point, Bob extracts from $(\beta ,\bar{%
\alpha})$ the actual value of the\emph{\ test variable} $\tau :=\beta -\bar{%
\alpha}$ which is then used to infer the total noise $\Delta _{B}=1+\sigma
^{2}$ affecting the signal. As soon as they recognize a non-tolerable noise,
i.e., $\sigma ^{2}>\tilde{\sigma}^{2}$ for some threshold noise $\tilde{%
\sigma}^{2}$, they stop the communication. Hereafter, we assume a
zero-tolerance protocol where no added noise is tolerated on the channel,
i.e., $\tilde{\sigma}^{2}=0$. We shall see that the QDC protocol can be
applied in realistic situations even with such a strict condition \cite{Zero}%
.

Let us show how the real-time check works in detail. For each control mode,
Bob collects the two quadratures $x=q,p$ of the test variable $\tau $. Then,
after $M$ control modes, he has collected $2M$ quadratures values $%
\{x_{1},x_{2},\cdots ,x_{2M-1},x_{2M}\}$ which give the estimator $%
v=\sum_{l=1}^{2M}x_{l}^{2}$. By using this estimator, Bob must distinguish
the two hypotheses%
\begin{equation}
H_{0}=\mathrm{no}\;\mathrm{Eve}\Leftrightarrow \sigma ^{2}=0~,~H_{1}=\mathrm{%
yes}\;\mathrm{Eve}\Leftrightarrow \sigma ^{2}\neq 0~.  \label{Hp_Test}
\end{equation}%
Let us fix the confidence level $r$ (i.e., the probability to reject $H_{0}$
though true) to a reasonably low value (e.g., $r=5\times 10^{-7}$). Then,
the hypothesis $H_{0}$ is accepted if and only if%
\begin{equation}
v<\mathcal{V}_{2M,1-r}~,  \label{AcceptsH0}
\end{equation}%
where $\mathcal{V}_{i,j}$ is the $j$th quantile of the $\chi ^{2}$
distribution with $i$ degrees of freedom. In other words, Alice and Bob
continue their direct communication in MM as long as the condition of Eq.~(%
\ref{AcceptsH0}) is satisfied in CM.

\begin{figure}[tbph]
\vspace{-0cm}
\par
\begin{center}
\includegraphics[width=0.45\textwidth] {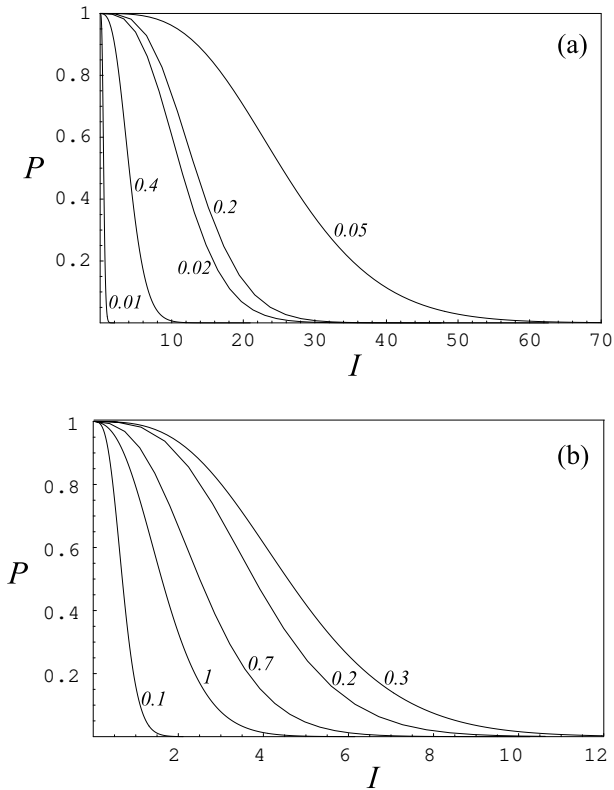}
\end{center}
\par
\vspace{-0.3cm}
\caption{Survival probability $P$ versus the number of stolen bits $I$. In
(a) no codes are used while in (b) a repetition code with $n=35$ is used.
The curves refer to UGQCM attacks with different values of added noise $%
\protect\sigma^2$. }
\label{doublepic}
\end{figure}

Let us explicitly analyze what happens when the channel is subject to
eavesdropping. In an individual UGQCM attack, Eve clones the signal input
and, then, heterodynes her output to estimate the signal amplitude $\bar{%
\alpha}$. After the release of the mask's value $\alpha _{M}$, Eve infers
the message amplitude $\alpha _{uu^{\prime }}$ and, therefore, the input
bits $(u,u^{\prime })$. In this process, Eve introduces an added noise $%
\sigma ^{2}$ on the Alice-Bob channel, while her output is affected by a
total noise equal to $\Delta _{E}=1+(4\sigma ^{2})^{-1}$. On the one hand,
we must compute the probability $\Pi _{M}(\sigma ^{2})$ that Eve evades $M$
control modes while introducing noise $\sigma ^{2}\neq 0$. After some
algebra we get
\begin{equation}
\Pi _{M}(\sigma ^{2})=\left[ \Gamma (M,0)-\Gamma \left( M,\frac{\mathcal{V}%
_{2M,1-r}}{2(1+\sigma ^{2})}\right) \right] \Big/(M-1)!~,  \label{Pi}
\end{equation}%
where $\Gamma (z,a):=\int_{a}^{+\infty }dt~t^{z-1}e^{-t}$ is the incomplete
gamma function. On the other hand, we must evaluate the amount of
information she can steal during her undetected life on the channel. Let us
assume that every input bit is a bit of information. As a consequence, the
stolen information per MM is equal to $I_{AE}=2[1-H(p)]$, where $%
H(p):=-p\log p-(1-p)\log (1-p)$ and $p=\varepsilon (\Omega ,\Delta _{E})$
can be computed from Eq.~(\ref{Intrinsic_error}). By combining $\Pi _{M}$
and $I_{AE}$, we can derive Eve's survival probability as a function of the
stolen information. Let $c$ be the probability of a control mode, so that $N$
runs of the protocol are composed by $cN$ control modes and $(1-c)N$ message
modes, on average. As a consequence, the survival probability will be $%
P:=\Pi _{cN}(\sigma ^{2})$ and the average number of stolen bits will be $%
I:=(1-c)NI_{AE}(\sigma ^{2})$. Then, for every value of $c$ and $\sigma ^{2}$%
, we can determine the function $P=P(I)$. Let us fix $c=69/70$ so that the
protocol has efficiency%
\begin{equation}
\mathcal{E}:=\frac{\#\mathrm{bits}}{\#\mathrm{systems}}=\frac{1}{35}~.
\end{equation}%
In Fig.~\ref{doublepic}, we have numerically plotted $P=P(I)$ for
several values of the added noise $\sigma ^{2}$. If the noise is
low, e.g., $\sigma ^{2}=0.01$, Eve steals very little information
($\simeq 1$ bit) while Alice and Bob complete an almost noiseless
QDC. In particular, Alice is able to transmit $\simeq 1.5\times
10^{4}$ bits of information by using $N\simeq 5\times 10^{5}$
systems. Notice that the maximum length of the QDC is roughly
bounded by the verification of $r^{-1}$ hypothesis tests and,
therefore, it is limited to about $4(1-c)(cr)^{-1}$ bits (i.e.,
$\simeq 1.2\times 10^{5}$ bits or $\simeq 4\times 10^{6}$ systems
using the above parameters). If the attack is more noisy (e.g.,
$\sigma ^{2}=1$), Eve again steals little information ($\simeq 1$
bit). In such a case, in fact, Eve is promptly detected by the
honest parties who, however, are prevented from exchanging
information (denial of service). According to
Fig.~\ref{doublepic}, Eve's best strategy corresponds to use a
UGQCM\ with $\sigma ^{2}\simeq 1/20$, so that she can steal
$\simeq 80$ bits before being revealed (for a cut off of $P=1\%$).
In such a case, Alice transmits $\simeq 630$ bits by using
$N\simeq 2.2\times 10^{4}$ systems.

How can we decrease the maximal amount of stolen information? One
possible solution is to increase further the control mode
probability $c$, so that the eventual presence of Eve is detected
before sending too many bits. However, this approach affects the
efficiency $\mathcal{E}$. An alternative and better solution
consists of making the decoding more sensitive to the presence of
added noise. Such an approach is possible by introducing classical
error correcting codes.

\section{Improving QDC via repetition codes}

In the basic scheme of QDC with continuous variables, noiseless
communication is possible up to an intrinsic error probability $\varepsilon $
depending on $\Omega $. In particular, such a probability decreases for
increasing $\Omega $. An alternative way for decreasing $\varepsilon $
consists of leaving $\Omega $ unchanged while introducing a classical error
correcting code. Such procedures are essentially equivalent for a noiseless
channel, since $\varepsilon $ is sufficiently small and the codes work very
well in that case. However, the scenario is different as the channel becomes
noisier. In such a case, in fact, the correcting codes have a non-linear
behavior which makes their performance rapidly deteriorate. Such a
non-linear effect can be exploited to critically split the correction
capabilities, and therefore the information gains, between the Alice-Bob
channel and the Alice-Eve channel.

\begin{figure}[tbph]
\vspace{-0cm}
\par
\begin{center}
\includegraphics[width=0.45\textwidth] {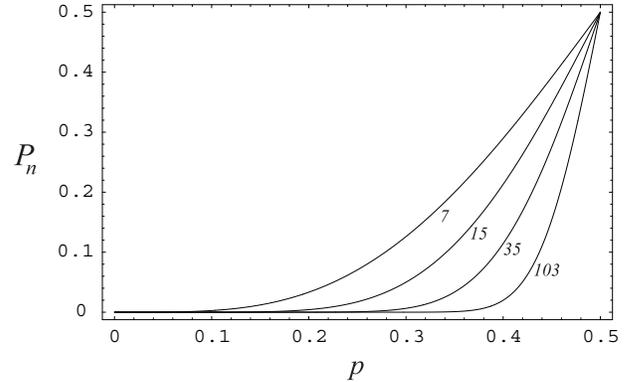}
\end{center}
\par
\vspace{-0.3cm}
\caption{Probability of an uncorrectable error $P_n$ versus the single
bit-flip probability $p$. Here, we consider repetition codes with $n = 7;\,
15;\, 35;\, 103$. }
\label{repetition}
\end{figure}

For the simplest case of an $n$-bit repetition code, an input bit $U=\{0,1\}$
is encoded into a logical bit $\bar{U}=\{\bar{0},\bar{1}\}$ of $n$ physical
bits via the codewords%
\begin{equation}
\bar{0}=\underbrace{00\cdots 0}_{n},\quad \bar{1}=\underbrace{11\cdots 1}%
_{n}.
\end{equation}%
By choosing an odd $n=2m+1$ (with $m=1,2,\cdots $), we can apply a
non-ambiguous majority voting criterion. This means that every bit-flip
error of weight $t<m+1$ is correctable, while every bit-flip error of weight
$t\geq m+1$ is not. Let us now consider a memoryless channel, where each
physical bit is perturbed independently with the same bit-flip probability $p
$ (as happens in the case of individual attacks). Then, the probability of
an uncorrectable error is given by%
\begin{equation}
P_{n}(p)=\sum_{k=m+1}^{n}\left(
\begin{array}{c}
n \\
k%
\end{array}%
\right) p^{k}(1-p)^{n-k}~.  \label{uncorregible_nbits}
\end{equation}%
For a sufficiently large $n$, the curve $P_{n}(p)$ displays a critical point
after which the correction capability suddenly starts to deteriorate very
quickly (see, e.g., Fig.~\ref{repetition}, showing $\tilde{p}\simeq 0.3$ for
$n=35$ and $\tilde{p}\simeq 0.4$ for $n=103$). Exactly these critical points
enable one to improve the QDC by transforming the communication protocol
into a threshold process, where the sensitivity to added noise is remarkably
amplified.

Let us choose a critical lattice's half-step $\tilde{\Omega}$, i.e., leading
to a critical intrinsic error probability $\varepsilon =\tilde{p}$. On the
one hand, when the channel is noiseless, Bob is able to recover the
codewords and reconstruct the logical bit with a very low error probability $%
P_{B}=P_{n}(\tilde{p})$ (that we call the \emph{logical} intrinsic error
probability). On the other hand, when the channel is noisy, Alice's
information is split into two sub-channels: the Alice-Bob channel, with
added noise $\sigma _{B}^{2}=\sigma ^{2}$, and the Alice-Eve channel, with
added noise $\sigma _{E}^{2}=(4\sigma ^{2})^{-1}$. The corresponding error
probabilities are respectively given by%
\begin{equation}
P_{B}=P_{n}(\tilde{p}+p_{B})~,~P_{E}=P_{n}(\tilde{p}+p_{E})~,
\end{equation}%
where $p_{B}=p_{B}(\sigma _{B}^{2})$ and $p_{E}=p_{E}(\sigma _{E}^{2})$ are
monotonic functions of the added noises (and are therefore linked by the
uncertainty principle). Now, if Eve tries to hide herself by perturbing the
Alice-Bob channel with a relatively small $p_{B}$, then her dual $p_{E}$
will always be big enough to perturb $\tilde{p}$ into the nonlinear region.
As a consequence, Eve will tend to experience $P_{E}\simeq 1/2$, gaining her
negligible information.

Let us explicitly show how to use an $n$-bit repetition code for
encoding/decoding. This is possible by simply adding pre-encoding and
post-decoding classical steps to the basic protocol. The message bits $%
(U,U^{\prime })$ are pre-encoded into a pair of logical bits%
\begin{equation}
\bar{U}=U_{1}U_{2}\cdots U_{n}~,~\bar{U}^{\prime }=U_{1}^{\prime
}U_{2}^{\prime }\cdots U_{n}^{\prime }~,
\end{equation}%
via the $n$-bit repetition code. Each pair of physical bits $%
(U_{k},U_{k}^{\prime })$ is then subject to the same encoding as before,
i.e., lattice encoding $(U_{k},U_{k}^{\prime })\rightarrow \alpha
_{u_{k}u_{k}^{\prime }}:=\alpha _{k}$, masking $\alpha _{k}\rightarrow
\alpha _{k}+\alpha _{M}=\bar{\alpha}$ and quantum preparation $\bar{\alpha}%
\rightarrow \left\vert \bar{\alpha}\right\rangle $. Then, after $n$ message
modes, Bob will have collected perturbed versions of the $n$ pairs $%
(U_{1},U_{1}^{\prime }),\cdots ,(U_{n},U_{n}^{\prime })$. By applying
standard error recovery (majority voting), he will then perform the
post-decoding of $(U,U^{\prime })$. In the same way as before, these
instances of MM (each one carrying a single physical bit of a codeword) must
be randomly switched with instances of CM, where Alice skips encoding and
simply sends Gaussian signals $\bar{\alpha}$ for testing the channel
(exactly as in Fig.~\ref{CMnoCodesPic})

Let us choose a repetition code with $n=35$. Then, consider a critical
half-step $\tilde{\Omega}=1$. Such a choice implies $\tilde{p}\simeq 32\%$
which leads to $\tilde{\varepsilon}\simeq 1\%$ for the logical bits $%
(U,U^{\prime })$. Then, let us also choose $c=1/2$, so that we
again achieve an efficiency $\mathcal{E}=1/35$. Let us then
analyze the effect of a UGQCM attack. On every cloned system (with
noise $\sigma _{E}^{2}$), Eve detects the complex amplitude via
heterodyne detection, therefore,
estimating Alice's signal amplitude $\bar{\alpha}$ up to a total noise $%
\Delta _{E}=1+\sigma _{E}^{2}$. After Alice's declaration of the mask $%
\alpha _{M}$, Eve derives the message amplitude and, therefore, a pair of
physical bits $(U_{k},U_{k}^{\prime })$. Each physical bit will be affected
by an error probability $p(\Delta _{E})$. After $n$ eavesdropped message
modes, Eve will be able to decode Alice's logical bit by majority voting up
to an error probability $P_{E}=P_{n}[p(\Delta _{E})]$. For each logical bit,
the acquired information is simply equal to $1-H(P_{E})$. As a consequence,
for each message mode, Eve acquires on average%
\begin{equation}
I_{AE}(\sigma ^{2})=2\left[ 1-H(P_{E})\right] /n~,  \label{Eve_stolen_codes}
\end{equation}%
bits of information (simply because $2$ logical bits are sent via $n$
physical systems).

Let us then consider the probability of Eve to evade $M$ control modes.
Since the control mode is implemented exactly as before, we have again $\Pi
_{M}(\sigma ^{2})$ as in Eq.~(\ref{Pi}). Such a quantity can be again
combined with the one of Eq.~(\ref{Eve_stolen_codes}). After $N$ runs of the
protocol, we have an average of $cN$ control modes and $(1-c)N$ message
modes, so that Eve's survival probability is again $\Pi _{cN}(\sigma
^{2}):=P $ and the stolen information is equal to $(1-c)NI_{AE}(\sigma
^{2}):=I$. Then, for every $\sigma ^{2}$, we can again evaluate the curve $%
P=P(I)$. According to Fig.~(\ref{doublepic}), the best choice for
Eve is a UGQCM\ with $\sigma ^{2}\simeq 0.3$, which enables her to
steal only $10$ bits of information before being detected (for
$P=1\%$). Such a result is a strong improvement with respect to
the basic protocol, where $80$ bits were left to Eve. Notice that,
for a low value of the noise like $\sigma ^{2}=0.1 $, Eve gets
$\simeq 1$ bit while Alice transmits $\simeq 320$ bits of
information by using $N\simeq 1.1\times 10^{4} $ systems. The
maximal length of QDC is here bounded by $4(1-c)(ncr)^{-1}\simeq
3500$ bits, i.e., $N$$\simeq 1.2\times 10^{5}$ quantum systems.

\section{Conclusion and discussion}

We have considered Alice and Bob confidentially communicating
without resorting to QKD. Such a task is in general risky and very
demanding. However, here we have shown how to construct a QDC
protocol which uses the same quantum resources as standard QKD,
even if they are exploited with a different logic. Such a protocol
is sufficiently confidential since it combines real-time checks of
the channel and a suitable masking of the secret\ information. In
particular, the maximum stolen information (i.e., the lack of
complete secrecy) can always be decreased by increasing the number
of controls at the expense of efficiency. As an alternative
approach we have also suggested the use of error correcting codes,
in such a way as to amplify the difference of information between
the eavesdropper and the honest user.

As a natural consequence of a demanding task like QDC, our
protocol allows an effective communication only when a small
amount of noise affects the quantum channel, thus restricting its
application to relatively short distances. Despite this
restriction, there are non-trivial situations where it can be used
in a profitable way. One of the possible applications concerns
entity authentication \cite{Book}, where one of the users (e.g.,
Bob) identifies the other (Alice) by comparing the bits of a
pre-distributed and secret \textit{authentication key}
$K_{\mathrm{aut}}$. Using the QDC protocol, the honest users have
the chance to perform this task without wasting too many quantum
resources. For instance, let us consider the case where Eve does
not perturb the quantum channel but impersonates Alice. Such a
quantum impersonation attack \cite{Dusek} is promptly revealed by
a small QDC session, where Bob receives directly the bits of
$K_{\mathrm{aut}}$ and, therefore, performs an immediate
comparison with his secret data. By contrast, in actual QKD\
protocols, such an attack can be revealed only after the
generation of the encryption key $K_{\mathrm{enc}}$ (to be used in
the private comparison of $K_{\mathrm{aut}}$). This clearly
requires the distribution and detection of many quantum states
(ideally infinite) and, therefore, the useless manipulation of a
huge amount of quantum resources (especially when entity
authentication is mutual). In general, since our QDC scheme adopts
the same quantum hardware as standard coherent state QKD, one can
also consider random switching between QDC (for authentication)
and QKD (for key generation).

As a final remark notice that our security analysis concerns the
case of individual attacks (where Eve does not exploit any quantum
memory). In future work it would be interesting to investigate the
performance of the protocol in the presence of collective attacks,
where Eve exploits a quantum memory to store all her output probes
and performs an optimal coherent measurement. It would be also
interesting to extend the security analysis to other forms of
Gaussian interactions (i.e., not referable to a UGQCM) and even to
non-Gaussian interactions, that may play a role against the usage
of repetition codes.

\section{Acknowledgements}

The research of S. Pirandola was supported by a Marie Curie
Outgoing International Fellowship within the 6th European
Community Framework Programme. S. Lloyd was supported by the W.M.
Keck center for extreme quantum information theory (xQIT).


\begin{thebibliography}{9}
\bibitem{Zero} A zero tolerance protocol does not promptly stop in realistic
situations (where $\sigma ^{2}\neq 0$) because the confidence level $r$
cannot be equal to zero.
\end{thebibliography}
\end{document}